\RequirePackage{lineno}
\documentclass[twocolumn,superscriptaddress,preprintnumbers,amsmath,amssymb,prc]{revtex4}  
\usepackage{graphicx}
\usepackage{dcolumn}
\usepackage{bm}
\usepackage{float}
\usepackage{color}
\usepackage{CJK}
\usepackage[colorlinks,linkcolor=blue,urlcolor=blue,citecolor=blue]{hyperref}
\usepackage{isotope}
\usepackage[normalem]{ulem}
\renewcommand{\sout}{\bgroup \color{red} \ULdepth=-0.5ex \ULset}

\usepackage{amsmath}
\usepackage{cases}
\usepackage{amssymb}
\usepackage{eqnalign}

\begin{document}
\begin{CJK*}{UTF8}{gbsn}
	
\title{The influence of hadronic rescatterings on the net-baryon number fluctuations}
\author{Qian Chen}
\affiliation{Guangxi Key Laboratory of Nuclear Physics and Technology, Guangxi Normal University, Guilin, 541004, China}
\affiliation{Key Laboratory of Nuclear Physics and Ion-beam Application (MOE), Institute of Modern Physics, Fudan University, Shanghai, 200433, China}
\affiliation{Shanghai Research Center for Theoretical Nuclear Physics, NSFC and Fudan University, Shanghai, 200438, China}

\author{Rui Wen}
\affiliation{School of Physics, Dalian University of Technology, Dalian 116024, China}
\affiliation{School of Nuclear Science and Technology, University of Chinese Academy of Sciences, Beijing, 100049, China}

\author{Shi Yin}
\affiliation{School of Physics, Dalian University of Technology, Dalian 116024, China}

\author{Wei-jie Fu}
\email[]{wjfu@dlut.edu.cn}
\affiliation{School of Physics, Dalian University of Technology, Dalian 116024, China}
\affiliation{Shanghai Research Center for Theoretical Nuclear Physics, NSFC and Fudan University, Shanghai, 200438, China}

\author{Zi-Wei Lin}
\email[]{LINZ@ecu.edu}
\affiliation{Department of Physics, East Carolina University, C-209 Howell Science Complex, Greenville, North Carolina 27858, USA}

\author{Guo-Liang Ma}
\email[]{glma@fudan.edu.cn}
\affiliation{Key Laboratory of Nuclear Physics and Ion-beam Application (MOE), Institute of Modern Physics, Fudan University, Shanghai, 200433, China}
\affiliation{Shanghai Research Center for Theoretical Nuclear Physics, NSFC and Fudan University, Shanghai, 200438, China}

	
\begin{abstract}
Fluctuations of conserved charges, such as the net-baryon number fluctuations, are influenced by different dynamical evolution processes. In this paper, we investigate the influence of hadronic rescatterings on different orders of cumulants of the net-baryon number distribution. At the start of hadronic rescatterings, we introduce net-baryon number distributions reconstructed based on net-baryon cumulants of different orders obtained from computation in functional renormalization group (FRG), where the distributions were constructed using the maximum entropy method. This way we introduce the critical fluctuations of Quantum Chromodynamics (QCD) into the AMPT model. Firstly, we find that hadronic rescatterings have distinct effects on cumulant ratios of different orders for the net-baryon number. Secondly, we observe that the effect of hadronic rescatterings is more significant for critical fluctuations than dynamical fluctuations, because the two-, three- and four-particle correlation functions due to critical fluctuations are weakened more significantly by hadronic rescatterings.

\end{abstract}
	
\maketitle

\section{Introduction}
\label{framework}

\label{introduction}

Investigating the phase structure of Quantum Chromodynamics (QCD) has been the focus of research in past, present, and future experimental heavy-ion collision programs~\cite{Bzdak:2019pkr, Luo:2020pef, ZY2023, KCM2023}. Because both the critical end point (CEP) and the first-order phase transition are associated with the characteristics of fluctuations, many experimental observables have been proposed to capture these features. The cumulants of conserved charges have been proposed as a promising observable~\cite{Stephanov:2008qz,Athanasiou:2010kw,Stephanov:2011pb,Cheng:2008zh,Gavai:2010zn,Chen:2021kjd}, as they are sensitive to the finer details of the phase structure in particular. In a simplified scenario of a system in equilibrium with volume $V$ at temperature $T$, the scaled susceptibilities of conserved charges of different orders ($\chi_{ijk}^{BQS}$) are defined as derivatives of the pressure with respect to the chemical potentials associated with the conserved charges (baryon number $B$, electric charge $Q$, and strangeness number $S$)
\begin{eqnarray}
\chi _{ijk}^{BQS}=\frac{\partial^{(i+j+k)}\left [ P/T^{4} \right ]}{{\partial(\mu _{B}/T)^{i}}{\partial(\mu _{Q}/T)^{j}}{\partial(\mu _{S}/T)^{k}}}=\frac{C_{ijk}^{BQS}}{VT^{3}},\label{MDIV1}	
\end{eqnarray}%
which determine the cumulants ($C_{ijk}^{BQS}$) of the distribution of the conserved charges in the grand canonical ensemble (GCE) for a system in thermal equilibrium. The cumulants of conserved charges are associated with powers of the correlation length, closely tied to susceptibility~\cite{Stephanov:2008qz}. They provide valuable insights into potential phase transitions, including remnants of chiral criticality at near-zero chemical potential~\cite{Friman:2011pf}.

From Eqs.~(\ref{MDIV1}), it is convenient for us to consider the ratios of cumulants, because they are intensive and volume-independent in the thermodynamic limit. The cumulant ratios are defined as follows~\cite{Ding:2015ona}:
\begin{eqnarray}
\frac{\chi _{2}}{\chi _{1}}=\frac{C_{2}}{C_{1}}=\frac{\sigma ^{2}}{M},\frac{\chi _{3}}{\chi _{2}}=\frac{C_{3}}{C_{2}}=S\sigma ,\frac{\chi _{4}}{\chi _{2}}=\frac{C_{4}}{C_{2}}=\kappa \sigma ^{2},\label{MDIV2}
\end{eqnarray}%
where the cumulants of event-by-event conserved charge multiplicity distributions are represented by $C_{n}$. In statistics, different cumulants characterize different features of a probability distribution. In the context of event-by-event particle multiplicity distribution, cumulants of the distribution play a crucial role, providing a basis for calculating correlation functions. The cumulants $C_{n}$ of multiplicity distributions of various orders can be calculated as~\cite{Luo:2011rg,Luo:2010by,Luo:2011ts,Luo:2017faz}:
\begin{eqnarray}
C_{1}&=&\left \langle N \right \rangle,\notag \\
C_{2}&=&\left \langle \left (\delta N  \right )^{2} \right \rangle,\notag \\
C_{3}&=&\left \langle \left (\delta N  \right )^{3} \right \rangle,\notag \\
C_{4}&=&\left \langle \left (\delta N  \right )^{4} \right \rangle - 3\left \langle \left (\delta N  \right )^{2} \right \rangle^{2},\label{MDIV3}
\end{eqnarray}%
where $N=N_{+}-N_{-}$ is the net number, and $N_{+}$ and $N_{-}$ denote the numbers of particles and antiparticles on an event-by-event basis, $\delta N=N-\langle N \rangle$ and $\left \langle \cdots \right \rangle$ represents an event average. 

Nonetheless, cumulants exhibit a limitation as they combine correlations of different orders. Therefore, a more insightful approach involves the examination of (integrated) multiparticle correlation functions~\cite{Bzdak:2019pkr,Bzdak:2016sxg,Kitazawa:2017ljq}. Utilizing the following relations allows for the investigation of integrated $n$-particle correlation functions $\kappa _{n}$, also referred to as factorial cumulants:
\begin{eqnarray}
\kappa _{1}&=&C_{1}=\left \langle N \right \rangle,\notag \\
\kappa _{2}&=&-C_{1}+C_{2},                         \notag \\
\kappa _{3}&=&2C_{1}-3C_{2}+C_{3},                   \notag \\
\kappa _{4}&=&-6C_{1}+11C_{2}-6C_{3}+C_{4}.           
\label{MDIV4}
\end{eqnarray}%

The phase structure of QCD matter, governed by strong interactions, can be unraveled through experiments involving heavy-ion collisions at various collision energies~\cite{STAR:2005gfr, STAR:2010mib, BRAHMS:2004adc, Friman:2011pf}. The extracted chemical freeze-out points from heavy-ion experiments span a range from nearly vanishing baryon chemical potential at the CERN Large Hadron Collider (LHC) to approximately $\mu_{B}\simeq 750$ MeV at the BNL Relativistic Heavy Ion Collider (RHIC). The QCD phase diagram has been extensively explored theoretically using various approaches, including lattice QCD from first principles~\cite{Ding:2015ona,HotQCD:2014kol,Bellwied:2013cta,Borsanyi:2014ewa,DHT2023}, functional renormalization group (FRG) in QCD~\cite{Fu:2019hdw,Fu:2022gou,YS2023}, effective theories~\cite{Schaefer:2004en,Herbst:2013ail,Herbst:2013ufa,Zhang:2017icm,Grossi:2021ksl,Chen:2021iuo,Otto:2022jzl,DYL2023}, and effective models~\cite{Fukushima:2003fw,Skokov:2010uh,Pisarski:2016ixt,Li:2017ple,XK2023}. A crucial question involves establishing connections between these quantities and experimental measurements. For instance, lattice QCD faces certain challenges in its calculations, with the emergence of the fermionic sign problem at finite baryon chemical potentials $\mu_{B}$~\cite{Bazavov:2012vg,Borsanyi:2013hza,Borsanyi:2014ewa,Ding:2015ona}. The functional renormalization group \citep{Mitter:2014wpa,Herbst:2013ufa,Fu:2019hdw,Fu:2021oaw} enables the study of equations of state at both high and low baryon chemical potentials. For small baryon chemical potentials, the functional renormalization group calculations yield a phase boundary curvature ($\kappa$) of $\kappa=0.0142(2)$ \cite{Fu:2019hdw}, closely aligning with recent lattice QCD computations, such as $\kappa=0.0145(25)$ in \cite{Bonati:2018nut}, $\kappa=0.0153(18)$ in \cite{Borsanyi:2020fev}, and $\kappa=0.0149(21)$ in \cite{Bellwied:2015rza}. Moreover, a recent estimate from FRG in QCD indicates that the location of CEP is in a small region around $(T_{CEP}$,$\mu _{B_{CEP}}$)$=$(107, 635) MeV. The FRG proves instrumental in describing critical properties at high baryon chemical potentials, offering valuable insights into locating the CEP in the QCD phase diagram.

The functional renormalization group calculations are confined to the hadronization stage and neglect interactions between hadrons and decay processes. In contrast, a multiphase transport model (AMPT) includes these dynamic processes. In addition, the FRG incorporates the mechanism of critical fluctuations, a feature not present in the AMPT model \cite{Chen:2022wkj,Chen:2022xpm,Huang:2021ihy,CQ2023}. The introduction of the FRG into the AMPT model simultaneously addresses both of these issues. Hence, in this study, we utilize the AMPT model to investigate the fluctuations of the net-baryon multiplicity distributions in Au+Au collisions at $\sqrt{s_{NN}} = 7.7$ GeV. Our main emphasis is on studying the influence of the hadronic rescatterings evolution stage on fluctuations in net-baryon multiplicity distributions. Specifically, we aim to understand how interactions between hadrons and decay processes in relativistic heavy-ion collisions modify critical fluctuations.

The paper's structure is outlined as follows. In Sec. II, we offer a brief introduction to the AMPT model and clarify how the FRG reconstructs net-baryon multiplicity distribution.  In Sec. III, we demonstrate the significant effects of hadronic interactions, and the critical fluctuation mechanism on the cumulants of net-baryon multiplicity distributions. Finally, Sec. IV concludes the paper.

\section{Model setup}
\label{framework}

\subsection{The AMPT model}
\label{ampt}
The AMPT (A Multi-Phase Transport) model with string melting mechanism, extensively utilized in investigating relativistic heavy-ion collisions~\cite{Lin:2004en,Lin:2021mdn}, encompasses four key components: initial conditions, parton cascade, hadronization, and hadronic rescatterings. The initial conditions establish the spatial and momentum distributions of minijet partons through QCD hard processes and soft string excitations, leveraging the HIJING model~\cite{Wang:1991hta,Gyulassy:1994ew}. The parton cascade delineates the evolution of partonic matter, featuring a quark-antiquark plasma resulting from the melting of excited strings. Parton scatterings are modeled by Zhang's parton cascade (ZPC)~\cite{Zhang:1997ej}, which currently incorporates two-body elastic parton scatterings using a perturbative QCD cross section (3 mb) with a screening mass. When all partons cease interactions, a spatial quark coalescence model combines nearby partons into hadrons. The resonance decays and hadronic reactions in the hadronic phase, involving baryon-baryon, baryon-meson, and meson-meson interactions, are described by an extended version of a relativistic transport model (ART)~\cite{Li:1995pra}. The AMPT model succeeds to describe multiple observables for relativistic heavy-ion collisions at the RHIC~\cite{Lin:2002gc,Lin:2004en} as well as the LHC~\cite{Lin:2021mdn,Ma:2016fve} energies, including HBT correlations~\cite{Lin:2002gc}, dihadron azimuthal correlation~\cite{Ma:2006fm,Wang:2019vhg}, collective flows~\cite{Bzdak:2014dia,Wang:2022rdh}, and strangeness production~\cite{Jin:2018lbk,Shao:2020sqr}. It has also been used to
study the chiral magnetic effects~\cite{Wang:2018ygc,Zhao:2019ybo,Liu:2020ymh,Wang:2021nvh}.
In this work, a new version of the AMPT model, which ensures the conservation of various charges (baryon number $B$, electric charge $Q$, and strangeness number $S$) in each hadronic reaction channel during the hadronic phase, is used to study the fluctuations of baryon numbers.

\subsection{functional renormalization group}
\label{sec:partB}

In the FRG approach, quantum and thermal fluctuations of different momentum modes are successively integrated in with the evolution of the renormalization group scale \cite{Wetterich:1992yh}, see Refs. \cite{Berges:2000ew, Pawlowski:2005xe, Braun:2011pp, Dupuis:2020fhh, Fu:2022gou} for reviews. The FRG is, therefore, well-suited for the studies of QCD phase transitions and QCD thermodynamics at finite temperature and densities, cf. \cite{Fu:2022gou} for a recent review. As we have discussed above, fluctuations of conserved charges, such as the baryon number fluctuations, are sensitive to the chiral phase transition as well as the critical dynamics pertinent to the CEP in the phase diagram. High-order cumulants of net-proton number distributions are utilized to search for the CEP in experiments, and recently a non-monotonic dependence of the kurtosis on the collision energy is observed with $3.1 \sigma$ significance by the STAR collaboration \cite{STAR:2020tga}.

\newcommand{\tabincell}[2]{\begin{tabular}{@{}#1@{}}#2\end{tabular}}
\begin{table}[htbp]
\centering
\caption{\label{tab:comparison} Baryon chemical potential $\mu_B$, pseudo-critical temperature for the chiral phase transition $T_{\mathrm{pc}}$, and volume of the fireball $V$ for different centrality bins in Au+Au collisions at $\sqrt{s_{NN}} = 7.7$ GeV, where the three parameters are used to reconstruct FRG net-baryon multiplicity probability distributions.}
\renewcommand\arraystretch{1.5} 
\scalebox{.75}{
\begin{tabular}{cccc}\\
\hline  \hline
Centrality &$\mu_{B}$ (MeV) &$T_{\mathrm{pc}}$ (MeV) & $V$ ($\mathrm{fm}^3$)\\
\hline
\tabincell{c}{0--5\%} & \tabincell{c}{399} &\tabincell{c}{139} &\tabincell{c}{980}\\
\hline
\tabincell{c}{5--10\%} & \tabincell{c}{395} &\tabincell{c}{140} &\tabincell{c}{701}\\
\hline
\tabincell{c}{10--20\%} & \tabincell{c}{391} &\tabincell{c}{140} &\tabincell{c}{508}\\
\hline
\tabincell{c}{20--30\%} & \tabincell{c}{382} &\tabincell{c}{141} &\tabincell{c}{340}\\
\hline
\tabincell{c}{30--40\%} & \tabincell{c}{376} &\tabincell{c}{143} &\tabincell{c}{212}\\
\hline
\tabincell{c}{40--60\%} & \tabincell{c}{357} &\tabincell{c}{144} &\tabincell{c}{126}\\
\hline
\tabincell{c}{60--80\%} & \tabincell{c}{337} &\tabincell{c}{146} &\tabincell{c}{48}\\
\hline  \hline
\end{tabular}
}
\label{TABLE}
\end{table}

In this work we utilize the cumulants of baryon number distributions calculated in a QCD-assisted low energy effective theory within the FRG approach in \cite{Fu:2021oaw, Fu:2023lcm}. Then, the baryon number distributions are reconstructed from the cumulants of different orders by means of the maximum entropy method or Gaussian process regression, which is detailed in \cite{Huang:2023ija}. In this work, the cumulants of the four orders are employed for the reconstruction of baryon number distributions. Values of the baryon chemical potential obtained in experiments\cite{STAR:2017sal} for corresponding to different centrality ranges at the collision energy $\sqrt{s_{NN}} = 7.7$ GeV are listed in Tab.~\ref{tab:comparison}. The temperature $T_{\mathrm{pc}}$ listed in the third column is the corresponding pseudo-critical temperature for every value of $\mu_B$, obtained in the calculations of FRG \cite{Fu:2021oaw, Fu:2023lcm}.

\subsection{The AMPT model with FRG sampling}
\label{ampt2}

\begin{figure}[htb]
\centering
\includegraphics[width=1.1\columnwidth]{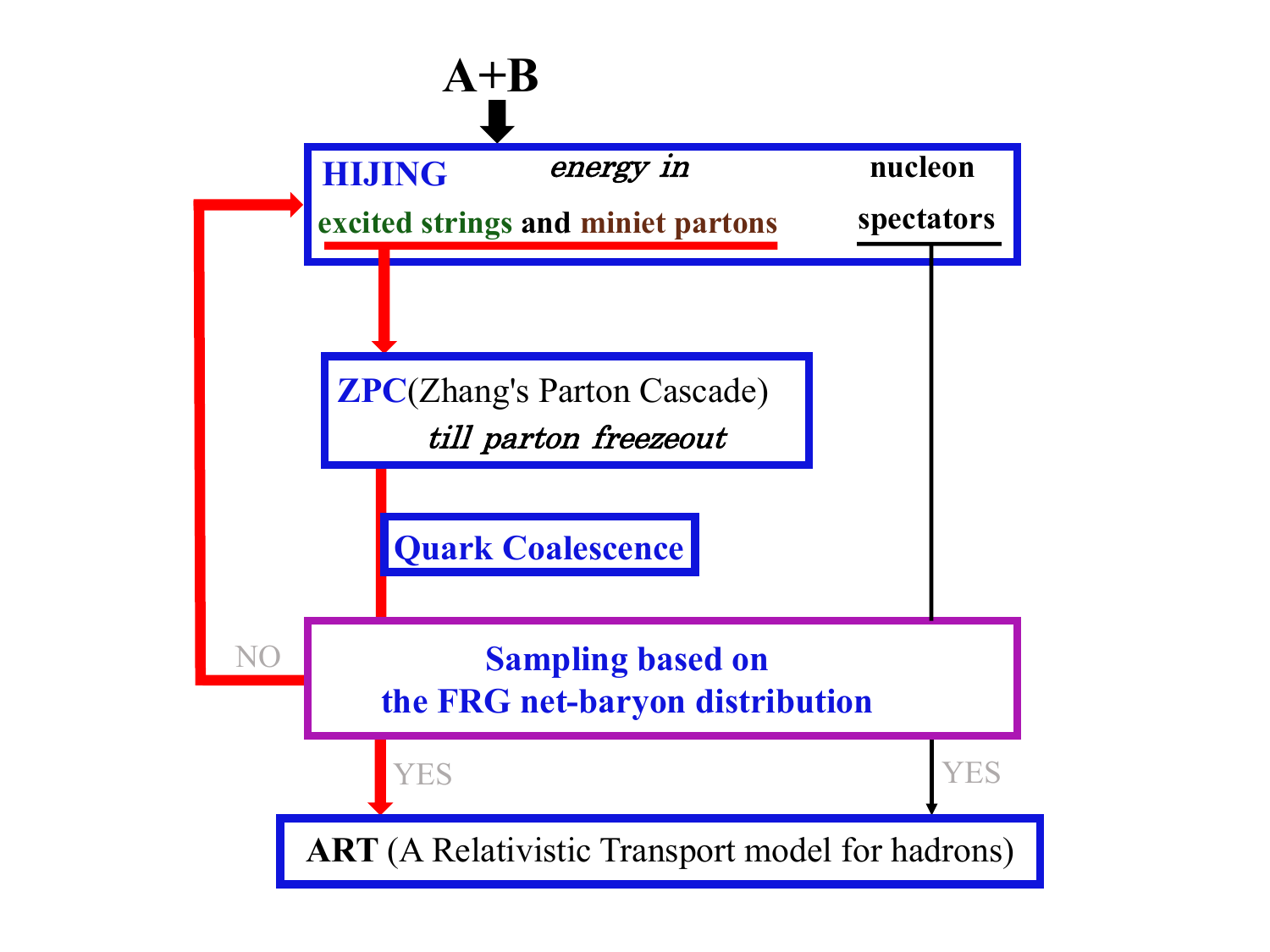}
\caption{(Color online) Structure of the AMPT model with string melting and FRG sampling.}
\label{FIG.1.}
\end{figure}

In order to study the influence of hadronic rescatterings on the critical fluctuations, we introduce the net-baryon probability distribution produced by the FRG to the AMPT model before hadronic rescatterings. As illustrated in Figure~\ref{FIG.1.}, we add the FRG sampling process to the AMPT model to introduce the critical fluctuations. Our method is to calculate the net-baryons produced by the AMPT hadronization in the chosen phase space and then select events so that they satisfy the net-baryon probability distribution of the FRG; the selected events then enter the next evolution stage of hadronic rescatterings.

\begin{figure}[htb]
\centering
\includegraphics[width=\columnwidth]{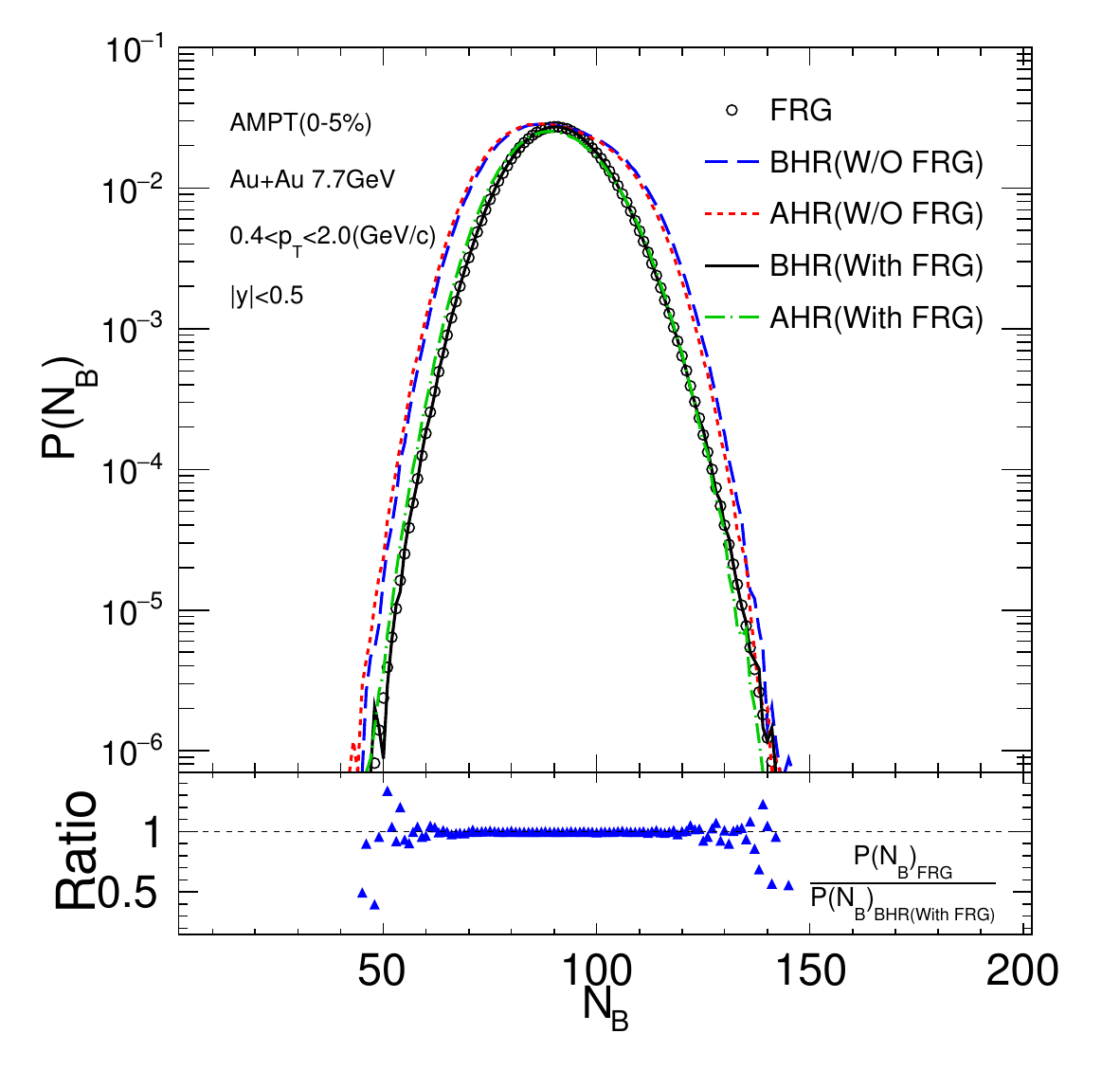}
\caption{(Color online)  Upper panel: The probability distributions of net-baryon multiplicity in  0--5\% central Au+Au collisions at $\sqrt{s_{NN}} = 7.7$ GeV. Lower panel: The ratio of the net-baryon multiplicity probability distribution from the FRG (circles) to the sampled one based on the FRG (solid curve) for the ``before hadronic rescatterings (BHR)" stage.}
\label{FIG.2.}
\end{figure}

 In our calculations, we apply the same kinematic cuts as used in the STAR experimental analysis~\cite{STAR:2021iop,STAR:2020tga} to calculate the aforementioned different cumulants of net-baryon multiplicity distributions. The unstable strange baryons (e.g., $\Lambda$, $\Sigma$, $\Xi$, and $\Omega$), which provide a feed-down contribution to protons and anti-protons, are also considered in our analysis. We select net-baryon within a transverse momentum range of $0.4<p_{T}<2.0$ GeV/$c$ and a midrapidity window of $\left |y   \right |< 0.5$. We use the charged particle multiplicity distribution to define centrality bins. To avoid self-correlation, we use the charged particle multiplicity other than protons and antiprotons within the pseudorapidity $\left |\eta \right |< 1$. We apply the $\Delta$ theoretical formula to estimate the statistical error calculation. More detailed information on the error calculations can be found in the references~\cite{Luo:2017faz,STAR:2021iop}.

\begin{figure*}[htbp]
\centering
\includegraphics[width=2.0\columnwidth]{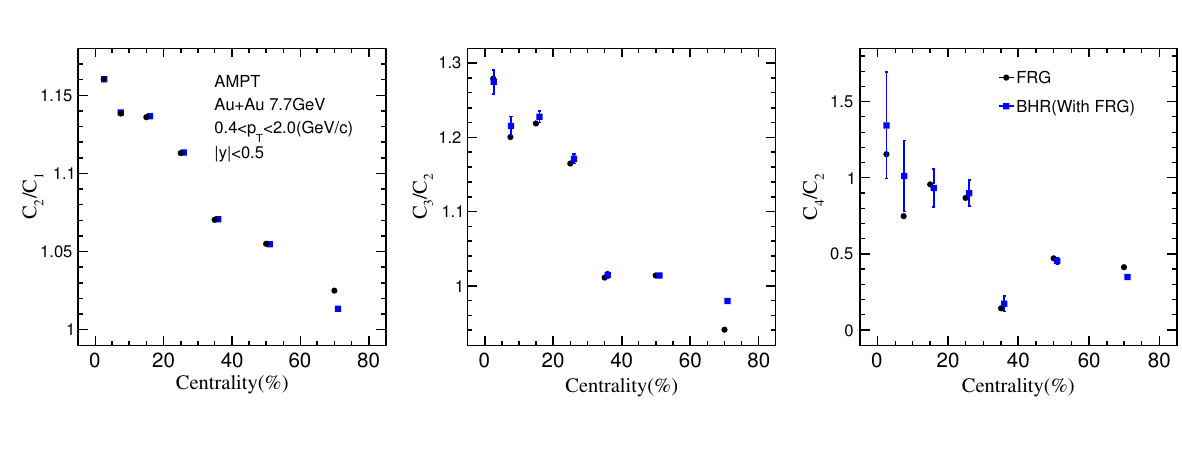}
\caption{(Color online)  The centrality dependences of the net-baryon cumulant ratios of $C_{2}/C_{1}$, $C_{3}/C_{2}$ and $C_{4}/C_{2}$ from the FRG, compared to the sampled one based on the FRG  for the ``before hadronic rescatterings (BHR)" stage.}
\label{FIG.3.}
\end{figure*}

Figure~\ref{FIG.2.} shows how we do the FRG sampling in detail. The upper panel displays the probability distributions of net-baryon multiplicity in the chosen phase space in  0--5\% central Au+Au collisions at $\sqrt{s_{NN}} = 7.7$ GeV, where the circles represent the net-baryon multiplicity probability distribution from the FRG method. The FRG distributions for different centrality bins are obtained according to the parameters in Table~\ref{TABLE}. As we can see, the net-baryon multiplicity distribution of ``before hadronic rescatterings'' from the AMPT model (dashed curve) [labeled ``BHR(W/O FRG)''] is very different from the FRG distribution (circles). In order to introduce the FRG distribution in the ``before hadronic rescatterings'' stage of the AMPT model, we sampled the net-baryon multiplicity distribution in this stage according to the FRG probability distribution. The sampled net-baryon multiplicity distribution based on the FRG for the ``before hadronic rescatterings'' stage (solid curve) [labeled ``BHR(With FRG)''] almost coincides with the FRG probability distribution. The lower panel in Fig.~\ref{FIG.2.} shows the ratio of the FRG net-baryon multiplicity probability distribution (circles in the upper panel) to the FRG sampled net-baryon multiplicity probability distribution (solid curve in the upper panel), where the ratio is mostly consistent with unity indicating success of our sampling. As shown in Fig.~\ref{FIG.1.}, all hadrons in the selected AMPT events that satisfy the FRG distribution participate in the following stage of hadronic rescatterings. We will compare the net-baryon distribution labeled ``after hadronic rescatterings (With FRG)''(dash-dotted curve) with the FRG distribution, as the difference between them reflects the effect of hadronic rescatterings on the imported net-baryon fluctuations. However, the difference between the ``after hadronic rescatterings'' stage based on FRG sampling (dash-dotted curve) [labeled ``AHR(With FRG)''] and the ``after hadronic rescatterings'' stage (dotted curve) without sampling [labeled  ``AHR(W/O  FRG)''] indicates the contribution the introduced critical fluctuations.

To further verify that the net-baryon multiplicity distribution of the ``before hadronic rescatterings'' stage based on FRG sampling essentially reproduces the FRG probability distribution, we calculate the net-baryon cumulant ratios in both cases. Figure~\ref{FIG.3.} shows the centrality dependences of the net-baryon cumulant ratios of $C_{2}/C_{1}$, $C_{3}/C_{2}$ and $C_{4}/C_{2}$ from the FRG (solid circles), compared to the sampled one based on the FRG  for the ``before hadronic rescatterings (BHR)" stage (solid square). We find these ratios from the FRG sampling are mostly in good agreement with the results of FRG within statistical errors, indicating that the sampling was done successfully.

From Table~\ref{TABLE}, FRG only provides the net baryon probability distributions for seven centrality bins, i.e. 0--5\%, 5--10\%, 10--20\%, 20--30\%, 30--40\%, 40--60\% and 60--80\%; therefore, AMPT results will be calculated for the seven centrality bins. However, the net proton measurements in the STAR experiment contain results for nine centrality bins in Au+Au collisions at $\sqrt{s_{NN}}=7.7$ GeV~\cite{STAR:2021iop}. To compare our results with the net proton STAR measurements~\cite {STAR:2021iop}, we applied the same centrality bin width correction (CBWC) to the following AMPT results as the STAR experiment in order to eliminate the effect of volume fluctuations~\cite{Luo:2011ts, Luo:2013bmi, Luo:2017faz, STAR:2021iop, Chen:2022wkj, Chen:2022xpm, Huang:2021ihy}.

\section{Results and Discussions}
\label{framework}

\begin{figure*}[htbp]
\centering
\includegraphics[width=2.0\columnwidth]{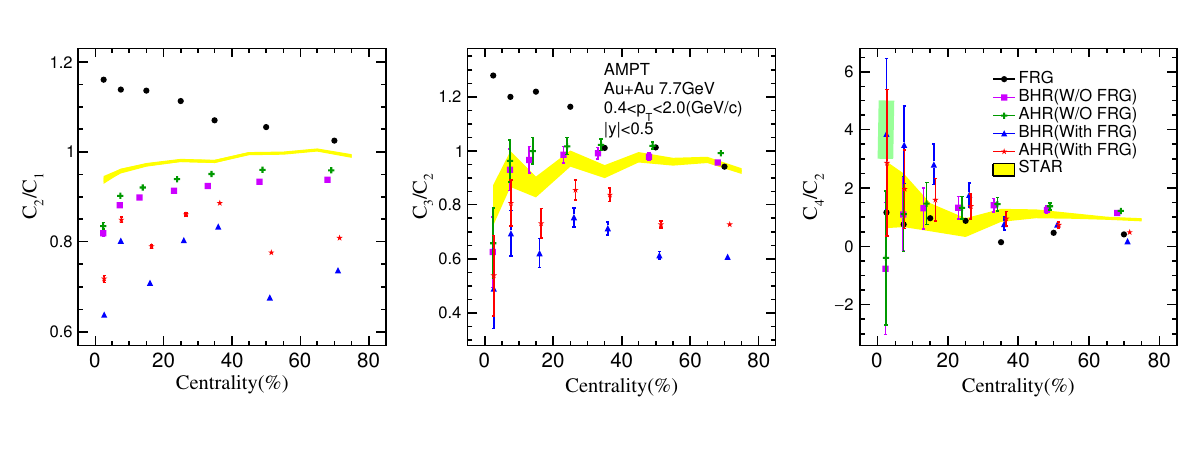}
\caption{(Color online) The centrality dependences of the net-baryon cumulant ratios of $C_{2}/C_{1}$, $C_{3}/C_{2}$ and $C_{4}/C_{2}$ in different stages in Au+Au collisions at $\sqrt{s_{NN}} = 7.7$ GeV,  compared to the net-proton STAR measurements~\cite{STAR:2021iop}. To facilitate plotting, the value of $C_{4}/C_{2}$ based on FRG sampling for the stage of ``before hadronic rescatterings (BHR)'' in the most centrality bin (0--5\%) (triangle symbols) has been scaled down by a factor of 2, indicated by the green box.}
\label{FIG.4.}
\end{figure*}

Figure~\ref{FIG.4.} shows the centrality dependences of the net-baryon cumulant ratios of $C_{2}/C_{1}$, $C_{3}/C_{2}$ and $C_{4}/C_{2}$ before and after hadronic rescatterings in Au+Au collisions at $\sqrt{s_{NN}} = 7.7$ GeV, compared to the net-proton STAR measurements~\cite{STAR:2021iop}. Without the FRG sampling, we observe that the net-baryon cumulant ratios $C_{2}/C_{1}$ and $C_{3}/C_{2}$ for the ``after hadronic rescatterings" stage (cross symbols) are larger than the results for the ``before hadronic rescatterings" stage (square symbols). Similarly, we can also observe that the net-baryon cumulant ratios $C_{2}/C_{1}$ and $C_{3}/C_{2}$ for the ``after hadronic rescatterings" stage based on FRG sampling (star symbols) are larger than the results for the ``before hadronic rescatterings" stage based on FRG sampling (triangle symbols). This suggests that the process of hadronic rescatterings increases the two cumulant ratios. We find that the cumulant ratios $C_{2}/C_{1}$ and $C_{3}/C_{2}$ based on FRG sampling for the ``before hadronic rescatterings'' and ``after hadronic rescatterings'' stages are smaller than the corresponding ratios in the AMPT model without the FRG sampling. Although there are relatively large errors, the cumulant ratios $C_{4}/C_{2}$ based on FRG sampling for the stages of ``after hadronic rescatterings'' and ``before hadronic rescatterings'' in the centrality bins of 0--5\%, 5--10\%, 10--20\% and 20--30\% seem to be larger than the corresponding ratios in the AMPT model without the FRG sampling. However, for the cumulant ratio $C_{4}/C_{2}$ for the other centrality bins, the ordering is reversed. All the differences indicate the impact of critical fluctuations on these cumulant ratios.

As shown in Fig.~\ref{FIG.3.}, we see that the cumulant ratios $C_{2}/C_{1}$,$C_{3}/C_{2}$ and $C_{4}/C_{2}$ are the same for the FRG and ``before hadronic rescatterings" with the FRG sampling. However, Fig.~\ref{FIG.4.} shows that the cumulant ratios for the FRG are noticeably larger than those for the ``before hadronic rescatterings'' with the FRG sampling because we have applied the CBWC for the correction of volume fluctuations. This suggests that volume fluctuations have an important impact on the cumulant ratios, which is consistent with previous studies~\cite{Chen:2022wkj,Chen:2022xpm,Huang:2021ihy}. We observe that the cumulant ratios $C_{2}/C_{1}$ and $C_{3}/C_{2}$ from the FRG significantly overestimate the STAR measurements and display a different dependence on the centrality. This is expected since the FRG calculations are done on the phase boundary as shown in Table~\ref{TABLE}, rather than on the chemical freeze-out. Incorporating the FRG into the ``before hadronic rescatterings'' stage of the AMPT model, our results underestimate the STAR measurements but the centrality dependence is consistent with the experimental measurements. Through the hadron rescattering evolution stage, the values increase but still underestimate the STAR measurements. The cumulant ratio $C_{4}/C_{2}$ from the FRG is consistent with STAR measurements. After incorporating the FRG into the ``before hadronic rescatterings'' stage of the AMPT model, we show that the value of $C_{4}/C_{2}$ overestimates the STAR measurements in central collisions but underestimates them in peripheral collisions. After hadronic rescatterings, the $C_{4}/C_{2}$ value decreases for mid-central to central collisions but increases for the most peripheral bin, bringing the model results with FRG closer to the STAR measurements.

\begin{figure*}[htbp]
\centering
\includegraphics[width=2.0\columnwidth]{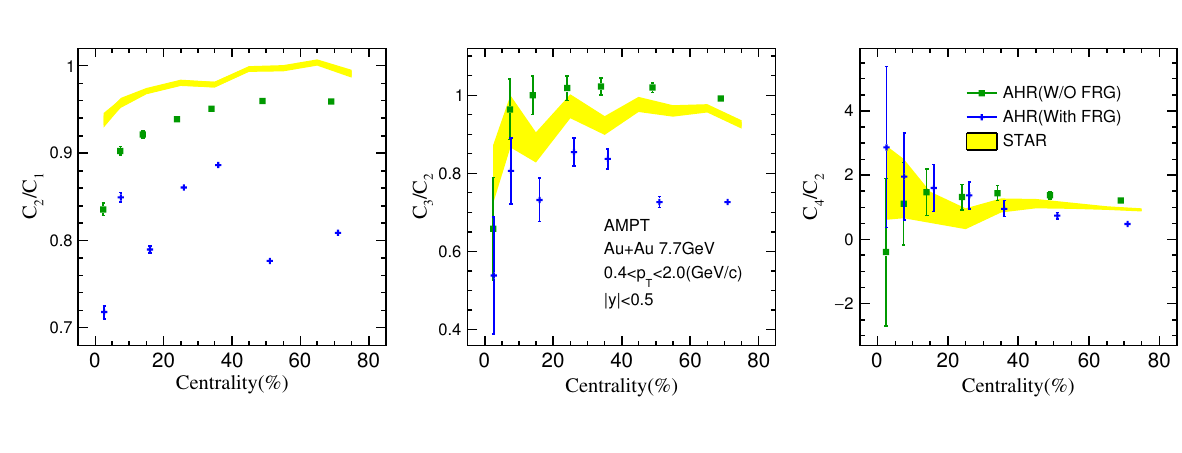}
\caption{(Color online) The centrality dependences of the net-baryon cumulant ratios of $C_{2}/C_{1}$, $C_{3}/C_{2}$ and $C_{4}/C_{2}$ with and without the FRG sampling in the ``after hadronic rescatterings (AHR)'' stage of Au+Au collisions at $\sqrt{s_{NN}} = 7.7$ GeV, compared to the net-proton STAR measurements~\cite{STAR:2021iop}.}
\label{FIG.5.}
\end{figure*}

To study the effect of critical fluctuations on the net-baryon fluctuations, we compare the AMPT results with and without the FRG sampling after hadronic rescatterings. Figure~\ref{FIG.5.} shows the centrality dependences of the net-baryon cumulant ratios of $C_{2}/C_{1}$, $C_{3}/C_{2}$ and $C_{4}/C_{2}$ with and without the FRG sampling in the ``after hadronic rescatterings (AHR)'' stage of Au+Au collisions at $\sqrt{s_{NN}} = 7.7$ GeV, where the original AMPT model (square symbols) contains dynamical correlations without critical fluctuations but the AMPT model with the FRG sampling (cross symbols) has both critical fluctuations and dynamical correlations. The results for the cumulant ratio of $C_{2}/C_{1}$ in both cases show a similar trend as the experimental measurements, albeit with lower values. In both cases, the cumulant ratio of $C_{3}/C_{2}$ follows the same trend as the experimental measurements. The AMPT model without the FRG sampling roughly agrees with the experimental measurements, while the AMPT model with the FRG sampling yields slightly lower values. Overall, critical fluctuations reduce both ratios. The trend of $C_{4}/C_{2}$ in the original AMPT model is different from the experimental measurement, although the statistical errors are large. In contrast, the trend in the AMPT with the FRG sampling looks more consistent with the experimental measurement, because the inclusion of critical fluctuations increases the ratio in central collisions and decreases the ratio in peripheral collisions. 

\begin{figure*}[htbp]
\centering
\includegraphics[width=2.0\columnwidth]{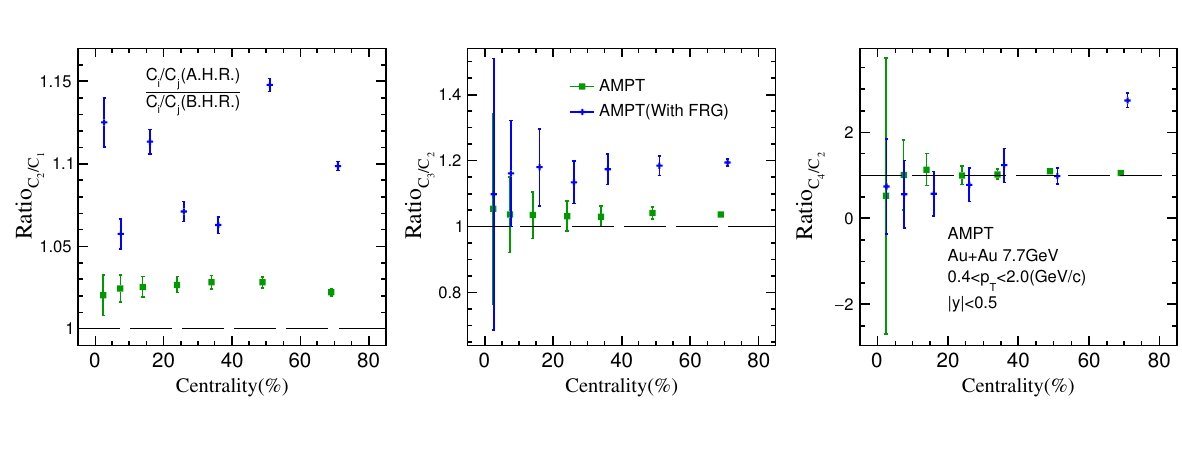}
\caption{(Color online) The net-baryon cumulant double ratio $\rm Ratio_{C_{i}/C_{j}}$ of the ``after hadronic rescatterings (AHR)" stage ${C_{i}/C_{j}(AHR)}$ to the ``before hadronic rescatterings (BHR)" stage ${C_{i}/C_{j}(BHR)}$.}
\label{FIG.6.}
\end{figure*}

To see the different effects of hadronic rescatterings in the two cases, Fig.~\ref{FIG.6.} shows the double ratio $\rm Ratio_{C_{i}/C_{j}}$ of the ``after hadronic rescatterings (AHR)" stage $\rm {C_{i}/C_{j}(AHR)}$ to the ``before hadronic rescatterings (BHR)" stage $\rm {C_{i}/C_{j}(BHR)}$. By comparing the double ratios of the AMPT model with and without the FRG sampling, we observe the effects of hadronic rescatterings on the cumulant ratios for the two different cases. Both $\rm Ratio_{C_{2}/C_{1}}$ and $\rm Ratio_{C_{3}/C_{2}}$ are larger than unity, which indicates that the effect of hadronic rescatterings increases the cumulant ratios of $C_{2}/C_{1}$ and $C_{3}/C_{2}$. This is consistent with our previous AMPT study on the effect of hadronic rescatterings on dynamical fluctuations~\cite{Chen:2022wkj}. Furthermore, the double ratios for the AMPT model with the FRG sampling are larger than those for the AMPT model without the FRG sampling, which indicates that hadronic rescatterings more significantly enhance $C_{2}/C_{1}$ and $C_{3}/C_{2}$ caused by critical fluctuations. In terms of the double ratios of $C_{4}/C_{2}$, the AMPT model without the FRG sampling shows results consistent with unity; however, the AMPT model with the FRG sampling shows a value smaller than unity in central collisions but larger than unity in peripheral collisions. It indicates that hadronic rescatterings decrease the cumulant ratio of $C_{4}/C_{2}$ in central collisions and increases the ratio in peripheral collisions for critical fluctuations.

\begin{figure*}[htbp]
\centering
\includegraphics[width=2.0\columnwidth]{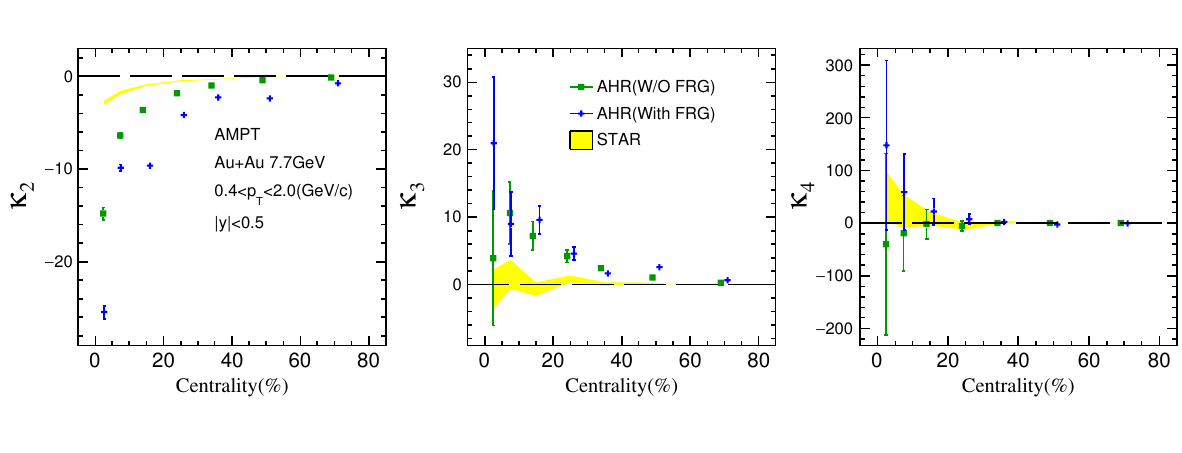}
\caption{(Color online) The centrality dependences of the net-baryon correlation functions of $\kappa_{2}$, $\kappa_{3}$ and $\kappa_{4}$ in the ``after hadronic rescatterings (AHR)'' stage of Au+Au collisions at $\sqrt{s_{NN}} = 7.7$ GeV, compared to the net-proton STAR measurements~\cite{STAR:2021iop}.}
\label{FIG.7.}
\end{figure*}

 As suggested in Refs.~\cite{Bzdak:2019pkr}, multiparticle correlation functions are much cleaner than cumulant ratios.  Figure~\ref{FIG.7.} shows the centrality dependences of the net-baryon correlation functions of $\kappa_{2}$, $\kappa_{3}$ and $\kappa_{4}$ in the ``after hadronic rescatterings (AHR)'' stage of Au+Au collisions at $\sqrt{s_{NN}} = 7.7$ GeV, compared to the net-proton STAR measurements~\cite{STAR:2021iop}. As the yield of antibaryons is much lower than that of baryons, the fluctuations of baryons are expected to be similar to that of net baryons in Au+Au collisions at $\sqrt{s_{NN}} = 7.7$ GeV. Therefore, we calculated the correlation functions using the method described in Eq.~(\ref{MDIV4}). It can be observed that the correlation functions $\kappa_{2}$ and $\kappa_{3}$ follow the same trends as the experimental measurements. In addition, we find that the strengths of the correlation functions $\kappa_{2}$ and $\kappa_{3}$ in the AMPT model without the FRG sampling are smaller than those in the AMPT model with the FRG sampling. Although our results for $\kappa_{4}$ show large statistical errors in both cases, the trend for the AMPT model without the FRG sampling seems to be the opposite of the experimental measurement. However, the trend for the AMPT model with the FRG sampling is in agreement with the experimental measurement. This suggests that the inclusion of critical fluctuations will change the four-particle correlation from negative to positive, which would be more consistent with the current experimental measurement.

\begin{figure*}[htbp]
\centering
\includegraphics[width=2.0\columnwidth]{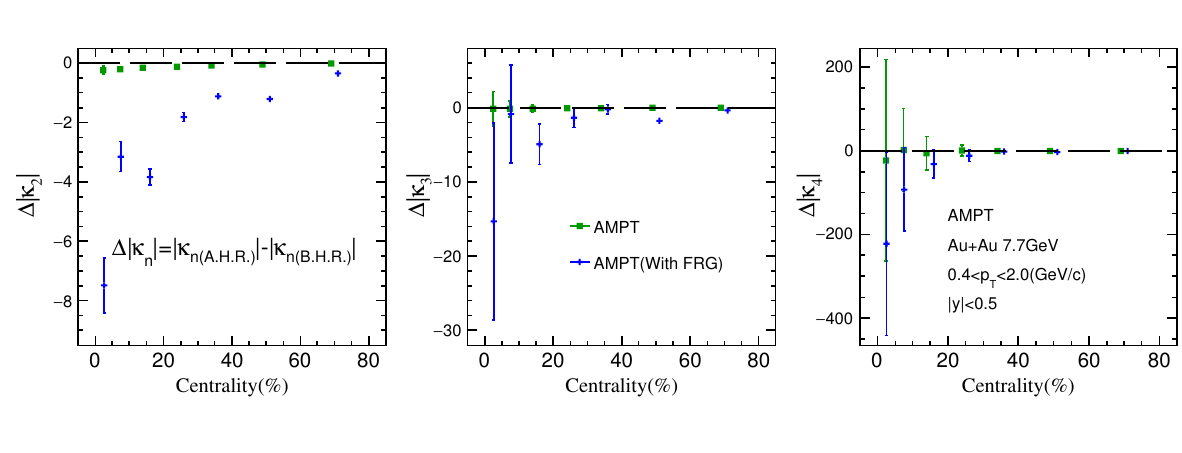}
\caption{(Color online) The centrality dependences of the difference $|\Delta\kappa_{n}|$ between the net-baryon correlation functions of the ``after hadronic rescatterings (AHR)'' stage $|\kappa_{n_(AHR)}|$ and that of the ``before hadronic rescatterings (BHR)" stage $|\kappa_{n_(BHR)}|$.}
\label{FIG.8.}
\end{figure*}

In our previous study~\cite{Chen:2022wkj}, we have found that multibaryon correlations are weakened after the evolution of heavy ion collisions. We next study the effect of hadronic rescatterings on the net-baryon correlation functions in the two cases. Figure~\ref{FIG.8.} shows the centrality dependences of the difference $\Delta|\kappa_{n}|$ of net-baryon correlation functions between the ``after hadronic rescatterings (AHR)'' stage $|\kappa_{n_(AHR)}|$ and the ``before hadronic rescatterings (BHR)" stage $|\kappa_{n_(BHR)}|$, i.e. $\Delta|\kappa_{n}|$ = $|\kappa_{n_(AHR)}|$ - $|\kappa_{n_(BHR)}|$. We find that the value of $\Delta|\kappa_{n}|$ is consistently less than zero for both cases, indicating that all two-, three- and four-particle correlation functions are weakened by hadronic rescatterings. We also observe that the values of $\Delta|\kappa_{n}|$ for the AMPT model with the FRG sampling are more negative than those of the AMPT model without the FRG sampling, which indicates that the critical fluctuation contributions to two-, three- and four-particle correlation functions are weakened more significantly by hadronic rescatterings. It should be noted that we focus on net-baryon fluctuations in this work, but previous work has shown that the behavior of net-proton fluctuations is similar to that of net-baryon fluctuations, with only differences in magnitude~\cite{Chen:2022wkj,Vovchenko:2021kxx}. 

\section{Summary}
\label{framework}

In summary, we incorporate critical fluctuations from the functional renormalization group (FRG) into the AMPT model in order to reveal the important effect of hadronic rescatterings on critical fluctuations. We observed apparent influences of hadronic rescatterings on the cumulant ratios of the conserved charge fluctuations. Hadronic rescatterings are found to increase the cumulant ratios $C_{2}/C_{1}$ and $C_{3}/C_{2}$ for all centrality bins but decrease $C_{4}/C_{2}$ in central collisions and increase $C_{4}/C_{2}$ in peripheral collisions. The effect of hadronic rescatterings is more significant for critical fluctuations than dynamical fluctuations. This is because the two-, three- and four-particle correlation functions due to critical fluctuations are weakened more significantly by hadronic rescatterings.

\section*{ACKNOWLEDGMENTS}
We thank Prof. Xiao-Feng Luo for helpful discussions, and Dr. Chen Zhong for maintaining the high-quality performance of the Fudan supercomputing platform for nuclear physics. This work is supported by the National Natural Science Foundation of China under Grants  No. 12325507, No. 12147101, No. 11890714, No. 11835002, No. 11961131011, No. 11421505, No. 12105054, No. 12175030, the Central Government Guides Local Scientific and Technological Development under Grant No. Guike ZY22096024, the National Key Research and Development Program of China under Grant No. 2022YFA1604900, the Strategic Priority Research Program of Chinese Academy of Sciences under Grant No. XDB34030000, the Guangdong Major Project of Basic and Applied Basic Research under Grant No. 2020B0301030008, and the National Science Foundation under Grant No. 2310021.

\bibliography{myref}


\end{CJK*}

\end{document}